\documentclass[twocolumn,floatfix,prl,aps,superscriptaddress,showpacs]{revtex4}
\usepackage{epsfig,graphicx,tabularx,amsmath}

\begin{document}
\title{Two-mode  effective  interaction in a double-well  condensate}
\author{D. M. Jezek}
\affiliation{IFIBA-CONICET, Pabell\'on 1, Ciudad Universitaria, 1428
  Buenos Aires, Argentina}
\author{P. Capuzzi}
\affiliation{IFIBA-CONICET, Pabell\'on 1, Ciudad Universitaria, 1428
  Buenos Aires, Argentina}
\affiliation{Departamento de F\'{\i}sica, FCEN, Universidad de Buenos Aires, Pabell\'on 1, Ciudad Universitaria, 1428
  Buenos Aires, Argentina}
\author{H. M. Cataldo}
\affiliation{IFIBA-CONICET, Pabell\'on 1, Ciudad Universitaria, 1428
  Buenos Aires, Argentina}

\date{\today}
\begin{abstract}
  We investigate the origin of a disagreement between the two-mode
  model and the exact Gross-Pitaevskii dynamics applied to double-well
  systems.  In general this model, even in its improved version,
  predicts a faster dynamics and underestimates the critical
  population imbalance separating Josephson and self-trapping regimes.
  We show that the source of this mismatch in the dynamics lies in the
  value of the on-site interaction energy parameter.  Using simplified
  Thomas-Fermi densities, we find that the on-site energy parameter
  exhibits a linear dependence on the population imbalance, which is
  also confirmed by Gross-Pitaevskii simulations.  When introducing
  this dependence in the two-mode equations of motion, we obtain
  a reduced interaction energy parameter which depends on the
  dimensionality of the system.  The use of this new parameter
  significantly heals the disagreement in the dynamics and also
  produces better estimates of the critical imbalance.
%
%
\end{abstract}
\pacs{03.75.Lm, 03.75.Hh, 03.75.Kk}

\maketitle

\textit{Introduction.}---Ultracold gases in double-well potentials exhibit a dynamics of matter
at its most basic quantum level when both wells couple via a
Bose-Josephson junction \cite{albiez05}. It is mostly interesting that
the slow passage from non-self-trapped to self-trapped states, which
is driven by a slow quench of the tunneling rate, is shown to obey the
Kibble-Zurek mechanism for a continuous quantum phase transition
\cite{Lee2009}. On the other hand, as a basic element of matter-wave
interferometers, such a double-well configuration represents a good
example of control of quantum coherence and entanglement in order to
achieve high-precision metrology devices \cite{Schumm2005}.  Being a
widespread ingredient for modelling in a diversity of areas, such as
quantum computing \cite{Strauch2008} and cosmology
\cite{Neuenhahn2012}, it is of fundamental importance to achieve the
best compromise between simplicity and accuracy in order to propose a
theoretical description for these systems.  In this respect, a
seemingly good balance between both requirements is represented by the
so-called two-mode (TM) model, which has been extensively studied in
the last years \cite{smerzi97,ragh99, anan06, jia08,
  albiez05,mele11,abad11}.  In particular, there is an active research
in double-well systems \cite{doublewell} which would greatly benefit
from an accurate TM model. The dynamics in the TM model rests on
assuming that the order parameter can be described as a superposition
of localized on-site wave functions with time dependent coefficients.
In 2005, both Josephson and self-trapping (ST) dynamics
\cite{smerzi97,ragh99} were experimentally observed by Albiez {\it
  et. al.}  \cite{albiez05} for large enough times so as to include
several oscillations.  They successfully measured the population
imbalance and the phase difference between sites during the
evolutions.  We note that the TM model has in general predicted a
sizable faster dynamics compared to both experiments and numerical
simulations of the Gross Pitaevskii (GP) equations
\cite{albiez05,mele11,abad11,gati2007}, in both Josephson and
self-trapping regimes.  The TM model predictions have also
systematically provided an underestimated value of the critical
imbalance for the transition between these regimes
\cite{albiez05,mele11,abad11,gati2007}.


The main purpose of this letter is to put forward an explanation for
such TM model disagreements. We revise the interaction term of the
Gross-Pitaevskii equation and find that the interaction effect is
overestimated by the TM model. Furthermore, for large number of
particles we calculate an effective interaction energy parameter which
reconciles the TM model with the numerical results.

{\it The two-mode model}.--- The TM model \cite{ragh99,anan06} assumes
the condensate wavefunction can be described as
\begin{equation}
\psi_{\mathrm{TM}}(r,t)= b_R(t)\, \psi_R({\bf r}) + b_L(t)\, \psi_L({\bf r}),
\label{varan2m}
\end{equation}
where $ \psi_R({\bf r})$ and $ \psi_L({\bf r})$ are real, normalized
to unity, localized on-site functions at the right and left wells,
respectively. The complex time-dependent coefficients $ b_R =
\sqrt{n_R} e^{i \phi_R}$ and $ b_R = \sqrt{n_R} e^{i \phi_R}$ verify $
n_R + n_L =1$ being $ n_k=N_k/N$, with $N_k$ the number of particles
in the $k$-site, and $N$ the total number of particles.  In order to
obtain the TM dynamics one introduces the order parameter into the
time-dependent GP equation,
\begin{equation}
i \hbar\frac{\partial\psi_{\mathrm{TM}}}{\partial t}=
\left[-\frac{ \hbar^2 }{2 m}{\bf \nabla}^2  +
V_{\rm{trap}}({\bf r})  +g\,N  \rho({\bf r},t)\right]\psi_{\mathrm{TM}} \, ,
\label{2t-dgp}
\end{equation}
where $m$ is the atom mass, $ V_{\rm{trap}}({\bf r}) $ is the
double-well potential, and $ g= 4 \pi a \hbar^2 / m $ is the coupling
constant with $a$ the scattering length.  The TM density $ \rho({\bf
  r},t)$ takes the form,
\begin{equation}
 \rho({\bf r},t)  =  n_R(t)  \rho_R({\bf r}) +  n_L(t)   \rho_L({\bf r} ) 
+ 2 \mathrm{Re} (b_R^* b_L)   \psi_R({\bf r})  \psi_L({\bf r}) ,
\label{rden}
\end{equation}
where we define the localized on-site densities $ \rho_k({\bf r}) = 
\psi_k^2({\bf r}) $ with $k=R,L$.  Multiplying Eq. (\ref{2t-dgp}) by
both $ \psi_R({\bf r})$ and $ \psi_L({\bf r})$ and integrating the
corresponding equations considering a symmetric double-well potential,
one obtains the equations of motion
\begin{align}
  i \hbar\,\frac{db_R}{dt}  =\, &\varepsilon b_R-Jb_L+U N|b_R|^2b_R
  + I N \left[(b_R^* b_L+b_R b_L^*)b_L \right. \nonumber \\ 
  &\left. + |b_L|^2 b_R \right] - F N
  \left[(b_R^*b_L+b_Rb_L^*)b_R
    + b_L\right],   \label{2model1} \\
  i \hbar\,\frac{db_L}{dt} =\, &\varepsilon b_L-Jb_R+U N |b_L|^2b_L
  +  I N \left[(b_L^* b_R + b_L b_R^*) b_R \right. \nonumber \\
  &\left.+| b_R|^2 b_L\right] 
  - F  N  \left[(b_L^*b_R+b_Lb_R^*)b_L + b_R \right] .
\label{2model2}
\end{align}
The on-site interaction energy parameter in $D$ dimensions is given by
\begin{equation}
U= g \int d^D{\bf r}\,\,  \rho_R^2({\bf r})=  g \int d^D{\bf r}\,\,  \rho^2_L({\bf r}) .
\label{U0}
\end{equation}
and the definition of the remaining parameters is given at the end of
the letter, see Eqs. (\ref{eps0})--(\ref{ijotap0}).  In the equations
of motion (\ref{2model1}) and (\ref{2model2}) all possible crossed
terms involved in the TM model have been considered, in accordance with
Ref. \cite{anan06}.  In terms of the population imbalance $ Z = n_R -
n_L $ and phase difference $ \phi= \phi_L- \phi_R$ the dynamical
equations can be rewritten as,
\begin{eqnarray}
 \dot{Z}  &  = & - \sqrt{1-Z^2}\sin\phi   + \gamma \,  (1 - Z^2) \sin 2 \phi \, , \\
 \dot{\phi} &  =  &\Lambda  Z  + \left[
\frac{Z}{\sqrt{1-Z^2}}\right]\cos\phi -   \gamma \,  Z (2+  \cos 2 \phi) 
\label{phasee}
\end{eqnarray}
where $ \Lambda = U N/{2 K} $, $ \gamma = { I N}/{2 K}$, and $ K = J +
F N $. The time derivatives have been expressed in units of $\hbar/2 K
$.

For the numerical simulations we consider a Bose-Einstein condensate
of Rubidium atoms confined by the external trap $ V_{\text{trap}} =
\frac{ 1 }{2 } \, m \, \omega ^2 {\bf r}^2 + V_b \, \exp ( -
x^2/\;\lambda_b^2)$.  We fix the trap frequency to $ \omega / (2 \pi)
= 77$ Hz, the number of particles to $ N=10^5 $, and for the barrier
we choose $ V_b = 50 \, \hbar \omega $ and a small width $ \lambda_b
=0.3 \, l $, with $ l = \sqrt{\hbar /( m \omega )}$ the harmonic
oscillator length.  In these conditions the TM parameters are $ U =
1.8993 \times 10^{-4} \, \hbar \omega $, $ J = 1.2915 \times 10^{-3}
\hbar \omega $, $ F = 2.0245 \times 10^{-8} \, \hbar \omega $, and $ I
= 3.996 \times 10^{-10} \, \hbar \omega $.

\begin{figure}
\includegraphics[width=0.9\columnwidth]{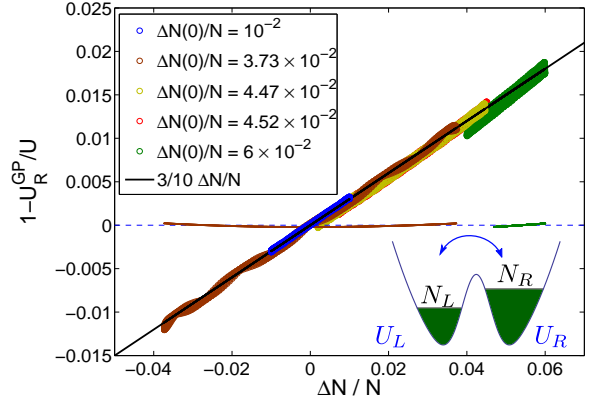}
\caption{(color online) GP simulations of $ 1-U^{\mathrm{GP}}_R(t)/ U
  $ (thick lines) and TM numerical calculations of $
  1-U^{\mathrm{\mathrm{TM}}}_R(t)/ U $ (thin lines) as functions of
  the population imbalance for different initial imbalances $
  Z(0)=\Delta N (0)/ N$.  For all the evolutions the initial phase
  difference is $ \phi(0)=0$.  As a solid black line we have also
  depicted $ 1-U_R/ U $ using $ U_R$ given by Eq. (\ref{Ue1du}). The
  inset shows a sketch of the double well potential with different
  populations. }
\label{fig1}
\end{figure}

{\it Modified model}.--- In order to identify the origin of the
disagreement in the dynamics within the TM model, we investigate the
effective potential felt by the system when the interaction term is
included.  To this aim, it is instructive to test the validity of the
TM model by numerically calculating
\begin{equation}
U^{\mathrm{GP}}_R(t) =  \frac{g}{n_R} \int d^D{\bf r}\,\,  \rho_{\mathrm{GP}}({\bf r},t)  \,  \rho_R({\bf r}),
\label{Ugp}
\end{equation}
as a function of time, where $ \rho_{\mathrm{GP}}({\bf r},t) $ is the
full GP density normalized to unity, and compare it with $
U^{\mathrm{\mathrm{TM}}}_R(t)= U + n_L(t) / n_R(t) I - 2 \mathrm{Re}
(b_R^* b_L)/ n_R(t) F $ predicted by the TM dynamics. This has been
obtained by replacing in (\ref{Ugp}) the GP density by the TM density
given by Eq. (\ref{rden}).  We note that
$U^{\mathrm{\mathrm{TM}}}_R(t) \simeq U$ within a $10^{-4}$ relative
error since the parameters $ F$ and $I$ are four orders of magnitude
smaller than $U$.  In Fig. \ref{fig1} we depict $
1-U^{\mathrm{\mathrm{TM}}}_R(t)/ U $ as a function of the population
imbalance for two initial $ Z(0)=\Delta N (0)/ N$ values. These curves
are almost vanishing within the error we have mentioned.
%
%
Instead of the vanishing value predicted by the TM model in
Fig. \ref{fig1}, the GP simulations of $ 1-U^{\mathrm{GP}}_R(t)/ U $
exhibit an almost linear behavior with the imbalance $\Delta N / N $.
The dispersion of the points is associated to the presence of sound
waves.

Aiming at reproducing this linear behavior within a simple approach,
we propose to take into account a more realistic density for
describing the interacting term of the GP equation. The idea is to
allow the localized on-site density to adopt the shape of what
corresponds to the true acquired population at a given time.  Thus, we
propose to replace in the density defined by Eq. (\ref{rden}) the
original $ \psi_R({\bf r}) $ and $ \psi_L({\bf r}) $ functions by
unbalanced localized on-site $ \psi^{\Delta N}_R({\bf r}) $ and $
\psi^{-\Delta N}_L({\bf r}) $, respectively, corresponding to
functions for artificial systems with a $ N \pm \Delta N $ total
number of particles with $ \Delta N(t) = N_R(t) - N_L(t) $.  These
wavefunctions are normalized to unity and will be referred to as
quasiestationary states, thus giving a density
\begin{multline}
  \rho^{\Delta N}({\bf r},t) \simeq n_R(t) \rho^{\Delta N(t)}_R({\bf
    r}) + n_L(t) \rho^{-\Delta N(t)}_L(\textbf{r})  \\ +
  \left[b_R^* b_L+b_R b_L^*\right] \psi^{\Delta N(t)}_R (\textbf{r})\psi^{-\Delta
    N(t)}_L(\textbf{r}),
\label{rdenm}
\end{multline}
with $\rho_{R,L}^{\pm \Delta N}$ the corresponding localized on-site
densities.

After performing such modifications the main correction is seen in
Eq. (\ref{phasee}), where the first term $ \Lambda Z $ changes to,
\begin{equation}
   N (  U_R   n_R  -  U_L    n_L )  \frac{  1}{ 2   K}  \, ,
\label{cambio}
\end{equation}
where $ U_R= g \int d^D{\bf r}\,\, \rho^{\Delta N}_R({\bf r})
\rho_R({\bf r}) $ and $ U_L= g \int d^D{\bf r}\,\, \rho^{-\Delta
  N}_L({\bf r}) \rho_L({\bf r} )$.  To estimate $U_R$ we resort to the
Thomas-Fermi (TF) approximation and assume that the barrier effect is
just to cut the condensate into two halves without modifying its
shape, note we have taken $ \lambda_b<< R_0$.  Thus, under these
assumptions, the localized on the right site density in $D=3$ for a
system with $ N + \Delta N $ particles may be approximated by
\begin{equation}
  \rho^{\Delta N}_R({\bf r})   
  \simeq  \frac{m\omega^2}{  g  ( N + \Delta N) }\left( R_{\Delta N}^2-r^2\right)  \Theta\left(R_{\Delta N} - r \right) \,  \Theta( x),
\label{roTF}
\end{equation}
 where $ \Theta( x)$ is the Heaviside function and  $  R_{\Delta N}$  denotes 
the corresponding TF radius.
Then  we may estimate,
\begin{equation}
U=   g \int d^3{\bf r}\,\,  \rho^2_R({\bf r})=  g 2 \pi  \int_0^{R_0} dr\, r^2 \,  
 \left[\frac{m\omega^2}{  g  N}( R_0^2-r^2)\right]^2.
\label{U03d}
\end{equation}
For evaluating $U_R$, we assume $\Delta N > 0$, in which case $ R_0 <
R_{\Delta N }$ because we are considering that the right site is more
populated than the ground state. Thus, the upper limit in the integral
in $r$ is $R_0$, giving
\begin{equation}
  U_{R}=     g 2 \pi  \int_0^{R_0} \!\!dr r^2 
  \frac{m\omega^2}{ g  N}( R_0^2-r^2)\,
 \frac{m\omega^2}{ g  (N+\Delta N)}(R_{\Delta N} ^2-r^2).
\label{U0R}
\end{equation}
After performing the integral and together with some algebra the
quotient of both magnitudes is
\begin{equation}
\frac{U_{R}}{U}= \left[\frac{7}{4} \left( \frac{R_{\Delta N}}{R_0} \right)^2 -  
 \frac{3}{4}\right]   \frac{N}{ N+\Delta N} .
\label{Ue0}
\end{equation}
Making use of the normalization condition for the density in a three
dimensional system, it is easy to verify that $ {R_{\Delta
    N}}/{R_0} = [{(N+\Delta N)}/ {N}]^{1/5}$, which inserted into
the previous formulae yields,
\begin{equation}
\frac{U_{R}}{U}= \left[  \frac{7}{4} \left(1 + \frac{\Delta N}{N}\right)^{2/5}-  \frac{3}{4}  \right] 
 \frac{1}{ 1 +\frac{ \Delta N}{N}}   \simeq  1  -   \frac{3}{10}   \frac{ \Delta N} {N},
\label{Ue1du} 
\end{equation}
to first order approximation in $ {\Delta N}/{N} $.  It is interesting
to note that the on-site parameter $U_R$ decreases when the population
on the site increases,  with respect to the stationary value, because the
new normalized on-site density ($ \rho^{\Delta N}_R({\bf r}) $)
spreads out over a wider region.

In Fig. \ref{fig1} we have depicted $1-U_R/U$ with $U_R$ given by
Eq. (\ref{Ue1du}).  Supposing that the density evolves in
quasiestationary conditions we may approximate,
\begin{align}
U^{\mathrm{GP}}_R(t)  &= \frac{g}{n_R} \int d^D{\bf r}\,\,   
 \rho_{\mathrm{GP}}({\bf r},t)  \rho_R({\bf r})  \nonumber 
\\ &\simeq  \frac{g}{n_R}  \int d^D{\bf r}\,\,  
 \rho^{\Delta N(t)}({\bf r},t) \, \rho_R({\bf r})  \nonumber \\
 &\simeq  g   \int d^D{\bf r}\,\,  
 \rho^{\Delta N(t)}_R({\bf r}) \rho_R({\bf r}) = U_R ,  
\label{UR}
\end{align}
which  is confirmed by the numerical results, as seen in Fig.\ \ref{fig1}.

On the other hand, proceeding in the same way on  the left site but
noting that in this case $ R_0 > R_{-\Delta N} $ one obtains,
\begin{equation}
  \frac{U_{L}}{U}=   \left[  \frac{7}{4} -  \frac{3}{4}   \left(1 -
      \frac{\Delta N}{N}\right)^{2/5}  \right] 
  \simeq  1  +   \frac{3}{10}   \frac{ \Delta N} {N}.
\label{Ue1}
\end{equation}
Then assuming the expressions Eqs.\ (\ref{Ue1du}) and (\ref{Ue1}) are
valid for every $ \Delta N(t)$ during the evolution, we introduce them
into the equation of motion for $ \dot{\phi} $, Eq. (\ref{phasee}),
and in particular the first term turns to (cf. Eq.\ (\ref{cambio})):
\begin{align}
\frac{U_{R}   N_R  -   U_{L}  N_L}{2K} &= \frac{U _R}{2K}
\left(\frac{ N + \Delta N}{2}\right)- \frac{U_L}{2K}
\left(\frac{ N-\Delta N}{2}\right) \nonumber \\
&=   \tilde{U}\,\frac{\Delta N}{2K}    
\label{resta}
\end{align}
with the effective interaction parameter $ \tilde{U}= \frac{7}{10} U
$. We note that the above approximation remains valid to second order
in the imbalance.  The same procedure can be applied in two and one
dimension. The results are summarized in Table I and it is
straightforward to prove that they are still valid for anisotropic
traps.  It is interesting to remark that even for very small
population imbalance, where the on-site interaction energies are almost
the same at the right and left sites ($U_R\simeq U_L \simeq U $), a sizeable
difference (a factor $7/10$) between them and the effective $
\tilde{U}$ parameter does exist.

\begin{table}
\caption{  On-site interaction energy corrections  for both right and
  left sites and the corresponding effective  interaction energy parameter $ \tilde{U}$ for systems in three, two and one dimensions.
}
\begin{ruledtabular}
\begin{tabular}{lccc}
Dimension  & $ U_{R}$ &   $ U_{L}$ &             $    \tilde{U} $    \\[3pt]
\hline \\[-5pt]
$ D=3$ &  $  (1  -   \frac{3}{10}   \frac{ \Delta N}{N} )U $ &   $  (1  +   \frac{3}{10}   \frac{ \Delta N} {N} )U $ &  $     \frac{7}{10}  U $ \\[3pt]
$ D=2$ &  $  (1  -   \frac{1}{4}   \frac{ \Delta N} {N} )U $ &    $  (1  +   \frac{1}{4}   \frac{ \Delta N} {N} )U $ &   $    \frac{3}{4}  U $ \\[3pt]
$ D=1$ &  $  (1  -   \frac{1}{6}   \frac{ \Delta N} {N} )U $ &    $  (1  +   \frac{1}{6}   \frac{ \Delta N} {N} )U $ &  $    \frac{5}{6}  U $ \\[3pt]
\end{tabular}
\end{ruledtabular}
\end{table}

\begin{figure}
\includegraphics[width=0.9\columnwidth]{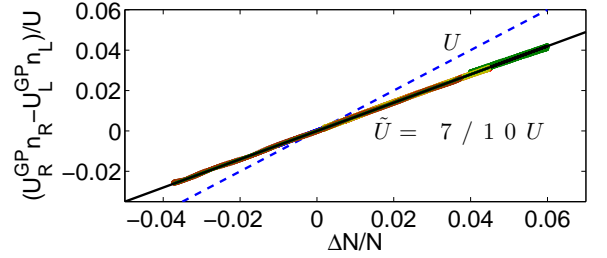}
\caption{(color online) GP simulations of $ ( U^{\mathrm{GP}}_R n_R -
  U^{\mathrm{GP}}_L n_L ) /U $ as a function of the population
  imbalance for the same initial conditions as Fig. \ref{fig1}.  The
  dashed line is the TM estimate while the solid one corresponds to
  the MTM model.  }
\label{fig2}
\end{figure}

In Fig.  \ref{fig2} we depict $ ( U^{\mathrm{GP}}_R n_R -
U^{\mathrm{GP}}_L n_L ) / U $ as a function of $ \Delta N /N $ for
several GP simulations.  We also plot $ ( U_R n_R - U_L n_L ) / U $
using Eq. (\ref{resta}) as a solid line, and the TM prediction as a
dashed line. It may be seen that the theoretical curve corresponding
to the effective interaction parameter $\tilde{U}$ much better
reproduces the numerical simulations.

On the other hand, we have also checked that when replacing
$\psi^{\Delta N}_{L,R}({\bf r})$ in the definitions of the parameters
$F$ and $I$, the corresponding numerical values change in less than
one percent with respect to the original ones.  

In view of the above results, we propose to modify the TM model by
replacing the on-site interaction energy parameter U by the new
constant parameter $ \tilde{U}$ in the original equation of motion
(\ref{phasee}). This will be called the modified two-mode (MTM)
model. In analogy to the TM model (disregarding the terms proportional
to $\gamma$), the new equations of motion can be derived from the
Hamiltonian,
%
\begin{equation}
 H^{\mathrm{MTM}} =\frac{ \tilde{\Lambda}}{2} Z^2   - \sqrt{1-Z^2}  \cos\phi  \, ,
\label{MTMham}
\end{equation}
where $ \tilde{\Lambda}= \tilde{U}  N / (2 K)$, and $Z$ and $\phi$ are
the canonical conjugate coordinates.

For our three dimensional system the renormalized parameters are $
\tilde{U} = 7/10 \times 1.8993 \times 10^{-4} \, \hbar \omega = 1.3295
\times 10^{-4} \, \hbar \omega $, and $ \tilde{\Lambda} = 2.004 \times
10^{3} \, \hbar \omega $.
\begin{figure}
\includegraphics[width=0.95\columnwidth]{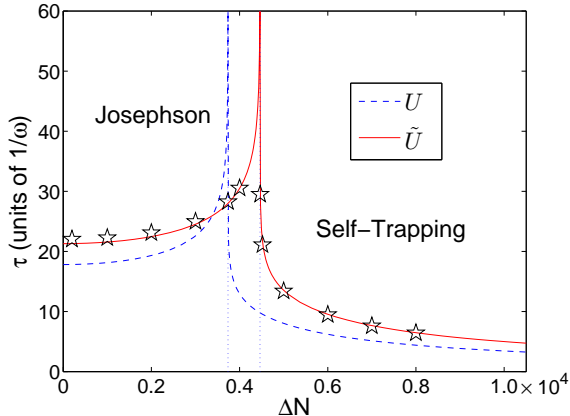}
\caption{(color online) The time period (in units of $1/\omega$) as a
  function of the initial $\Delta N$ is depicted using the TM model
  (dashed line in blue), the MTM model (solid line in red ), and GP
  simulations (stars).  }
\label{fig3}
\end{figure}
A very useful quantity for classifying the dynamics is the critical
imbalance ($ Z_c$), which separates initial conditions that evolve as
either Josephson or self-trapping orbits.  To appreciate one
consequence of our finding, we compare this quantity using the bare
on-site interaction parameter $\Delta N_c = N Z_c = 2 N /\sqrt{
  \Lambda} \simeq 3737 $, and the renormalized one, which gives a
larger estimate $\Delta N^R_c = 2 N / \sqrt{ \tilde{\Lambda} } =
\sqrt{10/7} \Delta N_c \simeq 4466$.  By means of numerical
simulations of the full GP equation, we have studied this transition
and checked it effectively occurs around $\Delta N^R_c$.

In Fig. \ref{fig3} we show the time period of the orbits using both
approximate approaches (TM and MTM) and the GP simulations. It may be
seen that the GP results are much better reproduced by the MTM
approach. This is particularly evident in the asymptotic values
corresponding to the critical imbalance where the period diverges. The
TM curves are in accordance with previous results where the
characteristic time was sizably underestimated with respect to both
experimental and GP results\cite{albiez05,mele11}. Only in a very
narrow interval around the critical imbalance, the relation between
the periods of the TM and the exact GP simulations is inverted, being
the TM time period larger.  On the other hand, the MTM model provides
a consistent agreement along the full range of imbalance studied.

\begin{figure}
\includegraphics[width=0.9\columnwidth]{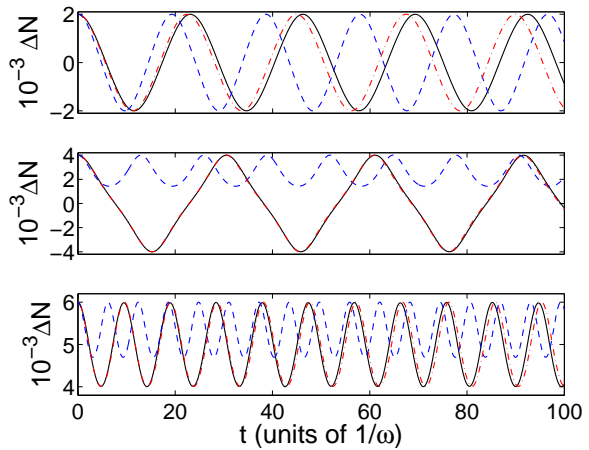}
\caption{(color online) Time evolutions of GP simulations (solid black
  line), TM model (dashed blue line), and MTM model (dot-dashed red
  line) for the initial conditions $ \Delta N(0) = 2000$ (top panel), $
  \Delta N(0) = 4000$ (middle panel), and $ \Delta N(0) = 6000$
  (bottom panel).  }
\label{fig4}
\end{figure}

To illustrate the different dynamics we show in Fig. \ref{fig4} the
difference of particles between the sites as a function of time for
different initial conditions.
It may be seen that the MTM model gives a much more accurate dynamics
than the TM model. Note that in the middle panel the MTM curve is
almost superposed with the GP one.

Finally, we now complete the definitions of the TM model  parameters,
\begin{equation}
\varepsilon = \int d^D{\bf r}\,\,  \psi_R({\bf r}) \left[
-\frac{ \hbar^2 }{2 m}{\bf \nabla}^2 +
V_{\rm{trap}}({\bf r})\right]  \psi_R({\bf r})
\label{eps0}
\end{equation}
\begin{equation}
J= -\int d^D{\bf r}\,\, \psi_R({\bf r}) \left[
-\frac{ \hbar^2 }{2 m}{\bf \nabla}^2  +
V_{\rm{trap}}({\bf r})\right]  \psi_L({\bf r})
\label{jota0}
\end{equation}
\begin{equation}
F= -  \,g\int d^D{\bf r}\,\,  \psi_R^3({\bf r})
 \psi_L({\bf r})
\label{jotap0}
\end{equation}
\begin{equation}
I= g  \int d^D{\bf r}\,\,   \psi_R^2({\bf r}) \,  \psi_L^2({\bf r}),
\label{ijotap0}
\end{equation}
where, due to the symmetry, we have chosen arbitrarily the right well
for performing the calculations.

\textit{Conclusions.}---By considering a more realistic effective
interaction we were able to construct a more accurate two-mode
model. Such a model is obtained by only replacing the on-site
interaction energy parameter by a reduced one, with simple scaling
factors depending on dimensionality. We have found that our modified
model reproduces much better the particle dynamics predicted by GP
simulations in a double well potential, both in the Josephson and
self-trapping regimes. 

This study can be extended to multiple well systems, in particular to
optical lattices with a high number of particles per site
\cite{anker05,jezek12}.

DMJ and HMC acknowledge CONICET for financial support under Grant No. 
PIP 11420090100243 and PIP 11420100100083, respectively. PC
acknowledges support from CONICET and UBA through Grants No. PIP 0546
and UBACYT 01/K156.

\end{document}